\documentclass{article}

\usepackage{arxiv}

\usepackage[utf8]{inputenc} % allow utf-8 input
\usepackage[T1]{fontenc}    % use 8-bit T1 fonts
\usepackage{hyperref}       % hyperlinks
\usepackage{url}            % simple URL typesetting
\usepackage{booktabs}       % professional-quality tables
\usepackage{amsfonts}       % blackboard math symbols
\usepackage{nicefrac}       % compact symbols for 1/2, etc.
\usepackage{microtype}      % microtypography
\usepackage{lipsum}
\usepackage{graphicx}
\graphicspath{ {./images/} }

\title{AI Chatbot for Generating Episodic Future Thinking (EFT) Cue Texts for Health
\thanks{Funding was provided through NIH NIDDK 3R01DK129567-02S1.}}

\author{
 Sareh Ahmadi \\
  Department of Computer Science\\
  Virginia Tech\\
  Blacksburg, VA 24061\\
  \texttt{saraahmadi@vt.edu} \\
   \And
 Edward A. Fox \\
  Department of Computer Science\\
 Virginia Tech\\
   Blacksburg, VA 24061\\
  \texttt{fox@vt.edu, 0000-0003-1447-6870} \\
}

\begin{document}
\maketitle
\begin{abstract}
We describe an AI-powered chatbot to aid with health improvement by generating 
Episodic Future Thinking (EFT) cue texts that should reduce delay discounting.
In prior studies, EFT has been shown to address maladaptive health behaviors.
Those studies involved participants, working with researchers, vividly imagining future events, and writing a description that they subsequently will frequently review, to ensure a shift from an inclination towards immediate rewards.
That should promote behavior change, aiding in health tasks such as treatment adherence and lifestyle modifications.
The AI chatbot is designed to guide users in generating personalized EFTs, automating the current labor-intensive interview-based process. 
This can enhance the efficiency of EFT interventions and make them more accessible, targeting specifically those with limited educational backgrounds or communication challenges. By leveraging AI for EFT intervention, we anticipate broadened access and improved health outcomes across diverse populations.
\end{abstract}

% keywords can be removed

\section{Introduction}
Conversational agents have exhibited significant potential in healthcare. 
AI-powered chatbots, which are a form of conversational agent, are dialogue systems designed to engage in natural language conversations with users to complete tasks. These agents are easily accessible via devices like smartphones and computers, making them particularly valuable for individuals with limited healthcare access or insufficient health knowledge. Conversational agents play a pivotal role in healthcare, serving two primary functions. First, they contribute to cost savings in healthcare by performing tasks such as health screening, guidance, home-based support, and medical staff training \cite{milne2020effectiveness}.  Their scalability and cost-effectiveness enable them to reach a wide population, thereby promoting patient-focused care and enhancing healthcare accessibility.

Second, conversational agents can be utilized by individuals to take control of their health. They can be applied for various health-related tasks, including behavior change, treatment assistance, and health tracking \cite{dingler2021use}. When these processes are automated, individuals can be more engaged, and healthcare professionals can be freed to focus on more intricate tasks, potentially improving healthcare quality for everyone \cite{milne2020effectiveness}.

Numerous studies have assessed the feasibility and effectiveness of AI chatbot interventions in promoting health behavior change. These studies have demonstrated the efficacy of AI chatbots in areas such as promoting healthy lifestyles, aiding smoking cessation, and ensuring treatment adherence \cite{aggarwal2023artificial}. These chatbots can leverage real-time user data to provide personalized services, making them a valuable tool for promoting and maintaining individual health. However, additional studies like this one are needed to confirm conclusions \cite{aggarwal2023artificial}.

In recent years, an increasing number of populations
have been targeted by Episodic Future Thinking (EFT) \cite{atance2001episodic} as an intervention to treat maladaptive health behaviors that may lead to lifestyle diseases \cite{bickel2023episodic, epstein2021delay, epstein2022imagine, epstein2022effects}. 

One of the aims is to decrease Delay Discounting (DD), which refers to the tendency to devalue delayed rewards in favor of immediate rewards. It has been shown that DD is associated with several lifestyle-related diseases \cite{bickel2012excessive}. When investigating the impact of EFT on both DD and health-related behavior, participants commonly produce written or spoken depictions of personally significant future events (e.g., weddings, birthdays, spending time with loved ones). These descriptions are subsequently utilized as prompts/cues to facilitate EFT during decision-making tasks that are conducted in a laboratory setting or in real-world environments \cite{sze2015web}. Gaining a better understanding of how EFT works, and the factors that influence its efficacy, is a focus of ongoing research \cite{sara}.
%\footnote{Funding was provided through NIH NIDDK 3R01DK129567-02S1}.

Interview-guided EFT cue generation is a method where researchers conduct interviews with participants to help them construct positive and realistic future event descriptions \cite{brown2022putting}. The goal is to create cues for EFT interventions, with each cue consisting of an introductory sentence and a detailed narrative description of the event (referred to as ``cues").
During the interview \cite{brown2022putting}, researchers ask participants to identify and describe positive future events that are plausible and could occur in the near future.
Participants are asked to list and describe future events that they are looking forward to.
These events must be vivid and easy for them to imagine. They will do this for different time periods. These events can be big or small. They can be related to health goals or unrelated.
To ensure the cues are personalized and vivid, researchers ask participants a series of questions about future events. Researchers may provide examples of good and bad cues, emphasizing the importance of including details and ensuring a positive emotional valence.  Participants rate events relative to a set of characteristics.  Researchers record participant responses verbatim, to minimize bias, and only use standardized questions to prompt additional details when necessary.

By analyzing the cue generation process, researchers can detect variations in cue quality between participants, and explore how cue quality may affect the effectiveness of EFT in future research \cite{brown2022putting}. This interview-guided method helps individuals construct future scenarios that are emotionally engaging and can be used for therapeutic purposes.

As a part of our work on expanding the impact of EFT, and guiding the work of others considering the use of AI chatbots to improve health, this paper makes two contributions:
\begin{enumerate}
    \item We describe and illustrate the operation of an AI chatbot that can enable a person interested in improving their health to generate Episodic Future Thinking (EFT) cues. This can serve as a valuable tool, particularly for individuals with lower levels of education, and those who lack writing skills or struggle with verbal communication, to assist them in the cue text generation process. It reduces the costs incurred with current human interview methods, ensuring scalability. The results should help expand the benefits of the EFT approach to promote improved health.
    \item We explain how LLMs, GPT-4, prompt tuning/engi-neering, moderation, chaining, memory, information extraction, and automated evaluation methods can be used to build and improve such chatbots.

\end{enumerate}
\section{Related Work}
\textbf{Chatbots:} The application of chatbots in healthcare has yielded significant advances across multiple domains. One notable area is in weight loss intervention \cite{chew2022use}. This research, offering personalized advice to users, underscores the effectiveness of AI chatbots when it comes to promoting balanced diets and physical activity. These chatbots, designed to mimic human interactions and operate on multiple platforms, enhance user engagement. However, it is essential to address challenges such as privacy concerns and interoperability to fully harness their potential in weight loss programs. Another study \cite{danieli2022assessing}  explores the impact of mental health applications, particularly the AI-powered agent TEO, on aging adults experiencing work-related stress. While the results did not demonstrate a significant advantage in using TEO alone or with cognitive behavioral therapy (CBT), a noticeable trend toward improving stress-related symptoms was observed within the treatment groups. Moreover, ChatGPT's performance in the United States Medical Licensing Exam \cite{kung2023performance}, showcases the potential of AI in medical education and clinical decision-making. However, there is need for further evaluation and comparison to ensure trust and explainability in clinical AI development. Accordingly, ethical considerations surrounding the use of Conversational Artificial Intelligence (CAI) in psychotherapy are explored  \cite{sedlakova2023conversational}, emphasizing the distinct role CAI should play in therapeutic contexts. 
Regarding under-served rural areas, an AI system acting as a ``virtual doctor"
 \cite{spanig2019virtual} could be particularly valuable.
 It could predict type 2 diabetes mellitus using non-invasive sensors, demonstrating the potential of AI in healthcare access and diagnosis.
Extending this approach, a chatbot can provide education and counseling to people with diabetes \cite{mash2022evaluating}.
This study showed that it was well-received, with a significant number of users engaging with it and reporting positive changes in self-management.
Other research \cite{nassar2023feasibility} explores the use of a diabetes chatbot to improve self-management in adults with type 2 diabetes. Participants engaged in weekly chats, including knowledge assessment and data reporting. Results showed high satisfaction, increased self-care confidence, and a drop in hemoglobin A1C levels for those who completed multiple chats -- highlighting the chatbot's potential to enhance diabetes care. Further research is needed to confirm these findings.

Putting all this in perspective, a systematic review and meta-analysis \cite{singh2023systematic} provides compelling evidence of chatbots' effectiveness in promoting improvements in physical activity, diet, and sleep. This study underscores the potential of chatbots to enhance overall well-being across various age groups and intervention durations, marking a significant advance in healthcare technology.

\textbf{LLMs, RLHF:} Traditional chatbot platforms primarily offer support for creating chatbots that follow predefined rules, and scripts that sound robotic and generate repetitive output \cite{adamopoulou2020chatbots}. Since the advent of deep learning, data-driven methods have gained attention for training Large Language Models (LLMs) using transformer architectures \cite{vaswani2017attention}, and their potential in developing human-like dialogue systems, such as Meena, \cite{adiwardana2020towards}, LaMDA \cite{thoppilan2022lamda}, and Dialogue-Prompted Gopher \cite{rae2021scaling}. Enhancing the conversational capabilities of these models for a broader range of topics and dynamic interactions requires extensive and domain-specific datasets. Even though these LLMs can perform multiple natural language processing tasks, they might produce unintended outputs like false information or biased content, often not adhering to user guidelines \cite{ouyang2022training}.
This misalignment stems from the primary goal of many LLMs, which is to predict the next word in publicly available content, rather than focusing on desired user-centric and safe responses \cite{ouyang2022training}. 

Given that LLMs are used widely in different natural language processing tasks, there is an effort to train them to align with user intentions. The ideal LLMs should be helpful, truthful, and avoid causing harm. As a result, a method was introduced to align language models with user goals across various tasks by refining them using Reinforcement Learning from Human Feedback (RLHF) \cite{christiano2017deep}.
One resulting model is InstructGPT \cite{ouyang2022training}.
RLHF is a complex process involving several stages, to create a conversational AI. Initially, a language model is pre-trained using traditional next-token or content prediction objectives. Then the model is further instruction-tuned on data where the output follows the input instructions. Next, a reward model is created to integrate human preferences into the system, using human rankings of model-generated text as input. The reward model aims to assign a scalar reward to text sequences, representing human preference based on standards such as being useful, truthful, and harmless. Finally, a reinforcement learning framework is applied to further tune the LLM to increase the probability of generating more highly rated outputs. 
RLHF has proven to be a highly successful method, garnering substantial attention from researchers and the AI community.

\textbf{ChatGPT, InstructGPT, GPT-4:} With the success of RLHF, ChatGPT \cite{WinNT} was built upon a comparable technique to InstructGPT, known for its impressive ability to generate high-quality responses.
Further development led to GPT-4 \cite{openai2023gpt4}.  
This transformer-based model benefited from post-training adjustments using RLHF, which led to enhanced accuracy and behavioral alignment, and significantly outperformed ChatGPT.
One study examines an early version of OpenAI's GPT-4 \cite{bubeck2023sparks}.
It argues that GPT-4, with its ability to handle tasks across various domains, nearing human-level performance, might represent an early form of artificial general intelligence (AGI). Considering the extensive and advanced abilities of GPT-4, its human-level performance, and its safety behavior improvement \cite{openai2023gpt4}, we chose this model to develop a chatbot tailored to our particular scenario.

\textbf{Prompt Tuning:} Prompt tuning is the dominant approach to get desired outputs from existing LLMs \cite{liu2023pre}. Prompts are textual instructions of the tasks, and examples of the desired outputs. Several studies have already shown the zero-shot and few-shot capability of LLMs \cite{brown2020language, weifinetuned, sanh2022multitask}. Zero-shot prompting involves presenting a language model with natural language queries or instructions without any examples. This approach relies on the model's ability to understand the inputs and respond accurately, using its built-in knowledge and language skills.
Few-shot prompting builds upon the zero-shot paradigm and focuses
on adapting language models to new tasks by providing the model with one or more examples of the desired output. Few-shot prompting has been shown to be effective in chat-based models \cite{madotto2021few, bae2022building}.
Creating strong prompts (i.e., prompt engineering) is essential to guide chatbots in performing specific tasks. Using prompts alone, developers can implement an LLM-powered chatbot  \cite{zamfirescu2023johnny}.
Aided by a suitable collection of specific prompts, LLMs can generate human-like conversation responses without the need for additional training data, effectively functioning as chatbots. In comparison to other frameworks, LLMs demonstrate significant potential in developing chatbots that are context-aware and capable of responding to off-topic users \cite{bae2022building}.  

Prompt tuning is an iterative process for building AI applications \cite{zamfirescu2023johnny} which includes the following steps \cite{book}.
\begin{itemize}
\item \textit{Initial Prompt Creation and Execution}: Create an initial prompt for the LLM based on the task. Execute the LLM with the initial prompt, knowing it may not produce the desired result.
\item \textit{Review and Analyze}: Evaluate the generated output to assess its quality and alignment with the goal.
\item \textit{Refine the Prompt}: Modify the prompt to provide clearer instructions, specifying details like word limits and formatting.
\item \textit{Iterative Process}: Repeat the process of refining and rerunning the prompt until the desired outcome is achieved.
\item \textit{Testing and Validation}: Continuously test and validate LLM responses against application requirements.
\end{itemize}

In essence, prompt engineering is an ongoing cycle of prompt development and refinement to guide the LLM. We employed an iterative process of prompt tuning to design an optimal template for our chatbot based on the needs and requirements of our application.

\textbf{LangChain:} One way to design an application is to use the LangChain library\footnote{https://www.langchain.com/}.
It is a novel framework designed for the development of LLM-powered systems.
LangChain is a library that offers a simple way to build application components and have effective memory management. Developing a chatbot serves as a prevalent use case for showcasing this functionality.
LangChain offers different types of memory.
Each takes the history of the conversation with the chatbot and formats it in a different way.
For example, since each chat model (LLM) has a certain context limit, if the conversation is long and reaches that limit, then with LangChain conversation summary memory\footnote{https://python.langchain.com/docs/modules/memory/types/summary\textunderscore buffer}, the conversation history can be stored in a summarized format.
The summarization can be performed by another LLM.
The resulting summary can be used for the current query (user request).
This approach can be used in case an interview session exceeds the LLM's context limit.

\textbf{OpenAI Evals8:} 
In order to find any shortcomings and continue to improve the quality of the chatbot, evaluating the outputs of an LLM is important; it is part of prompt engineering development. 
OpenAI Evals8\footnote{https://github.com/openai/evals} is a  framework designed to assess the capabilities of models such as GPT-4.
It can closely examine performance
\cite{openai2023gpt4}. Evals8 is fully compatible with established benchmarks and serves as a tool for monitoring model performance during deployment. Based on the provided documentation\footnote{https://github.com/openai/evals/blob/main/docs/eval-templates.md}, in situations where the expected model output can exhibit considerable diversity, such as responding to an open-ended question, it is discovered that employing the model to assess its own performance is a feasible approach for automated evaluation.

\section{Method}
The goal is to build a chatbot (AI assistant) that can guide people to write an effective and vivid EFT.
The AI-generated EFTs can be used by the participants to improve their health outcomes. 

As explained above, prior EFT studies investigated the impact of EFT on maladaptive health behaviors, to improve health-related outcomes.
Those efforts required interviewing and instructing the participants to generate a series of cues for the desired outputs.  
The participants were asked to write short texts (cues) about the events for different time frames. Participants generated detailed and vivid descriptions of these events. An example of selected data, from one participant in an earlier published study \cite{sara}, is as follows:
\begin{itemize}
\item In about 1 month, I am playing golf with my friends. We are having a great time and enjoying the company and competition. We laugh and have a great time.
 
\item  In about 3 months, I am picking my daughter up from college. I am excited she is done with school and we go to lunch at our favorite sushi restaurant and enjoy each other's company.
 
\item  In about 6 months, I am fishing in the bay with friends. We are on a charter boat and excited to catch some nice fish. We bet on who will catch the biggest fish.
%We bet on who will catch the biggest fish.
\end{itemize}
To guide the participants to have a detailed narrative description of the event, participants are guided to describe plausible positive events that might occur in the future. They begin with an introductory sentence for their event. To make the event feel real and immediate, participants are asked detailed questions about the event's circumstances, including their actions, companions, and personal feelings. Researchers provide examples of effective and less effective cues, highlighting the need for detailed, positively charged descriptions. Events should be concise with a clear start and end, typically lasting under a day. Participants then evaluate the event's emotional value and vividness.  Interviewers ensure participants adhere to guidelines, using the present tense and describing vivid events.

Our AI assistant system has been designed and built to guide the EFT cue generation process in an automated manner.

\begin{figure*}

\centerline{\includegraphics[width = 15cm]{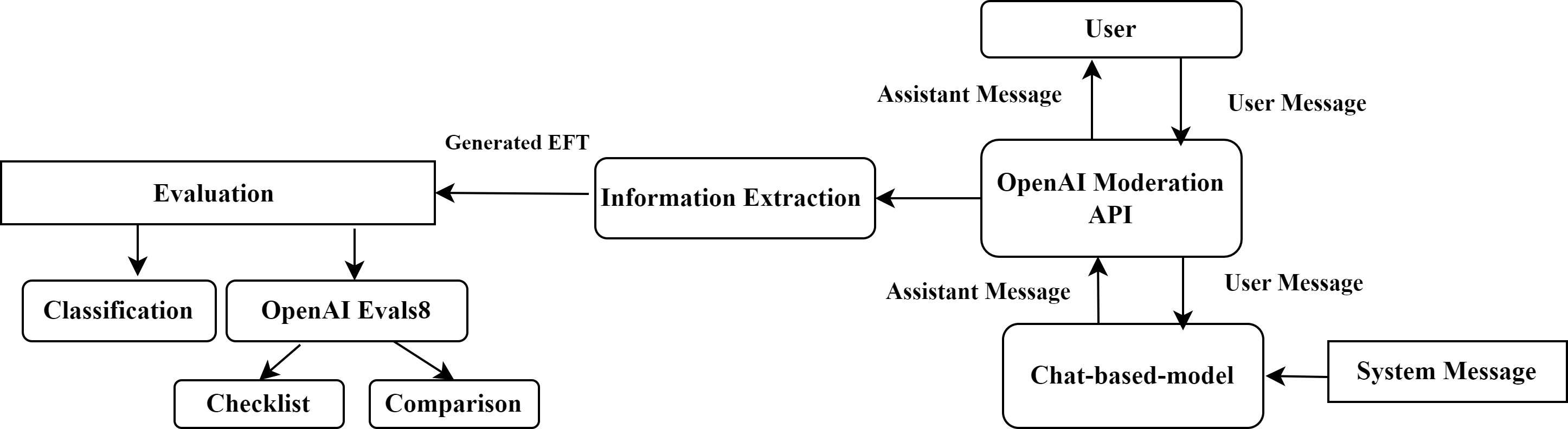}}
\caption{Chatbot (AI assistant) framework}
\label{fig1}
\end{figure*}

\textbf{System to Generate EFT Cues:} Fig. \ref{fig1} gives the architecture of our chatbot system. We explain the different modules below.

\textit{System Message:} One way to customize GPT-4, within certain limits, is to use a system message\footnote{https://openai.com/research/gpt-4}. The system message establishes the behavior or attitude we want the LLM to adopt. 
The term `Assistant Message' refers to a message generated by the chat-based AI language model, which has been created with the purpose of generating responses in accordance with the user's input.  For an input user message, the model will generate a suitable response in accordance with the request from the user message and in alignment with the overall behavior specified in the system message. The model analyzes the entire conversation history, which includes both user and assistant messages, in order to produce responses that are pertinent and logically connected.
%It considers the context and information available in the conversation history to produce responses that are precise and fitting.
The input system message in Fig. \ref{fig2} is designed to adapt the system to our scenario and to make it a customized chatbot, focused on EFT cue generation.

\begin{figure*}

""" 

You are an AI assistant to help people generate personalized Episodic Future Thinking (EFT) texts in which people imagine different events vividly for different time frames and write about them. 

You respond in a short, very conversational friendly style. 

If the user asks something not related to EFT, politely tell the user you can not help with that.

"""

\caption{Input system message}
\label{fig2}
\end{figure*}

\textit{Moderation:} A tool for checking the input user request and also the output of the model is the OpenAI Moderation API\footnote{https://platform.openai.com/docs/guides/moderation}. This is important to ensure the quality and safety of the system. Although as model quality improves, the likelihood of harmful outputs diminishes, it is still recommended to use this API \cite{openai2023gpt4} for building applications. The Moderation API helps developers identify and filter prohibited content across various categories. It helps verify that the system is being used in a responsible way, and can prevent any potential misuse,
%. It is also essential to verify the generated outputs by the chatbot, 
especially for sensitive audiences.  The Moderation API can be employed to filter and oversee content, and to take appropriate actions such as sharing canned fallback responses or generating new corrective responses.

\begin{figure*}

\centerline{\includegraphics{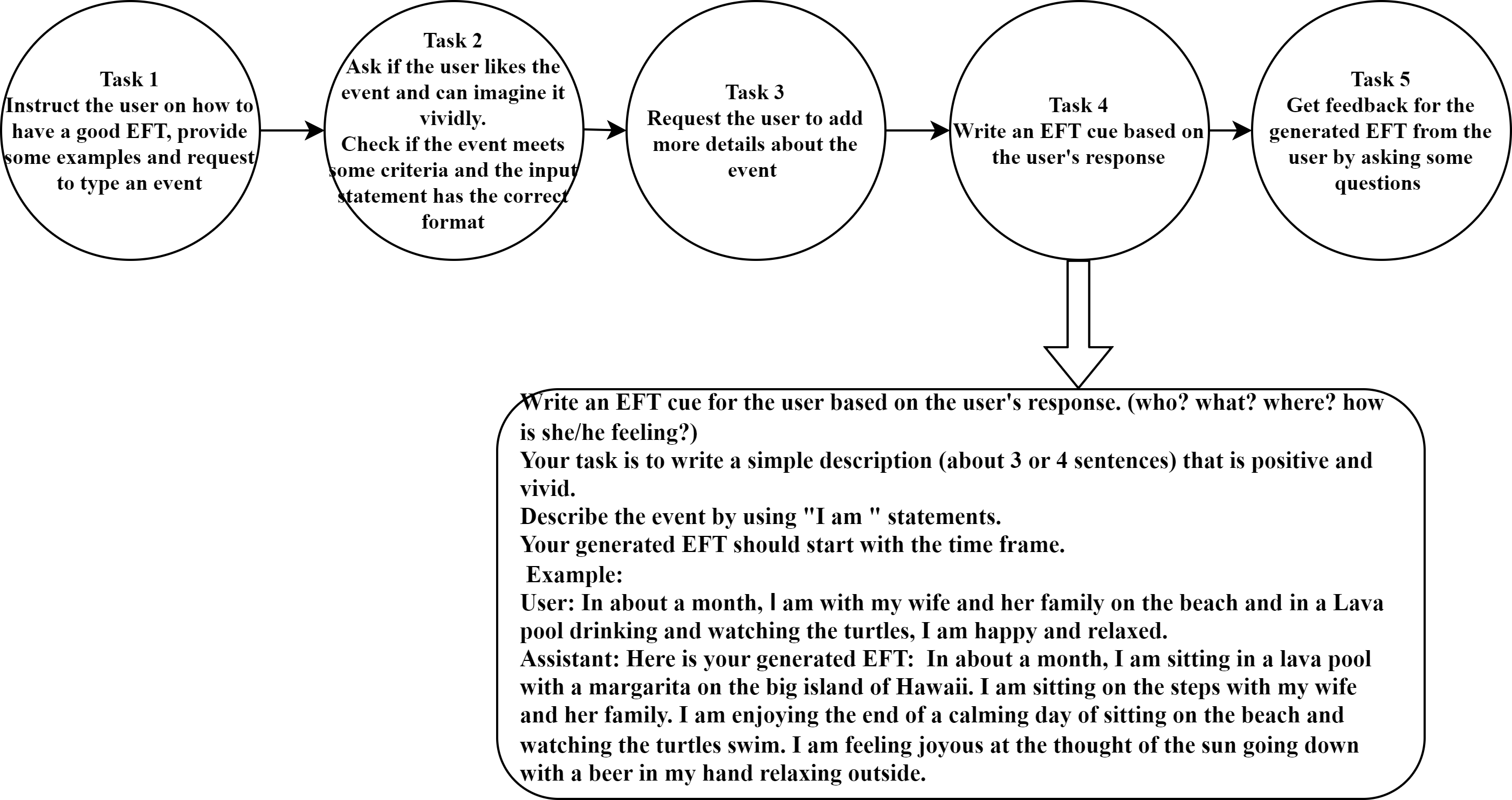}}
\caption{Sub-tasks for EFT generation}
\label{fig3}
\end{figure*}

\textit{Chaining:} Chaining is a technique to break complex tasks into smaller steps for LLMs, each with natural language prompts, thereby improving success rates \cite{wu2022ai}. The results obtained from one or more previous steps are combined and used as input for the subsequent steps. This approach allows users to apply the same model to multiple sub-tasks, increasing the likelihood of success for each sub-task compared to attempting to solve the entire problem in a single step. Fig. \ref{fig3} shows the task of generating an EFT cue that is explained earlier. The process is split into sub-tasks for the AI assistant to perform.
Each circle details a sub-task and has its own instruction to perform the sub-task. We have included textual instructions for Step 4, which prompts the model to generate an EFT based on the previous sub-tasks when the user is requested to add more details about the event, i.e., who the user is with, what the user is doing, where the user is, and how he/she is feeling. The prompt includes the instructions for the sub-task followed by an example of the desired output for the few-shot learning setting.

\textit{Memory:} LLM-based conversational agents respond to current queries based on previous conversations or interactions. By storing the previous interactions, the LLM is given a memory to remember the conversation history. Building upon this, we have integrated the memory feature into our system to store all the conversations. GPT-4 has a context limit (maximum number of tokens) of 8K tokens. Notably, for the scope of this particular interview format, a context limit of 8K tokens is adequate.

\textit{Information Extraction:} When the model generates the EFT for the user, it follows up with some questions to get feedback from the user. The generated personalized EFT and the user's feedback need to be saved for further study. The capabilities of GPT-4 in extracting information from texts have been shown \cite{liu2023summary}, with the potential of extracting different types of entities. We have added the information extraction module which basically prompts the language model (GPT-4) to extract the required information from the conversation stored in the memory and save it in JSON format. 
The module is capable of transforming text in JSON through prompt design. The prompt shown in Fig. \ref{fig4} can extract the generated EFT and the participants' scores from the conversation between the user and the chatbot.

\begin{figure*}
 
prompt = """

Create a JSON summary of the previous conversion. 

The fields should be:

1) generated EFT 

2)  Each question is the key and the value for each question should be the user's score.

"""   

\caption{Extracting generated EFT and user's feedback}
\label{fig4}
\end{figure*}

% For future interview formats, if other information and entities need to be extracted, suitable information extraction prompts can be added after each chain, in order to not lose any information for long conversations that need to be summarized. 

%\afterpage{

%}

\begin{figure*}[ht]
prompt =
    """
    
    You are evaluating a generated response based on the context that the agent uses to generate the response.
    
    ************
    
    [User]: In about 1 month, I am horseback riding with my best friend on a farm in the fall. I am very happy.
    
    ************
    
    [Assistant]: In about 1 month, I am horseback riding with my best friend on a beautiful farm. The fall leaves are painting the
    landscape with warm colors. I am feeling incredibly happy, the joy bubbling up inside me as we ride together, laughing and
    chatting. The feeling of the horse moving beneath me is wonderful. It’s a perfect day, and I am filled with a deep sense of
    contentment and joy.

    ************
    
    Compare the factual content of the generated response with the context. \
    Ignore any differences in style, grammar, or punctuation.
    Answer the following Yes/No questions:
    
    Question 1: Does the generated response describe a vivid event based on the definition of vividness? (Yes or No)

       Vivid definition: The text contains sufficient and specific details so that the event described is readily and easily imaginable. A considerable amount of context has been given regarding the event.
    
    Question 2: Does the generated response use the correct format (i.e. start with a time frame, for example, In about a month)? (Yes or No) 
    
    Question 3: Does the generated response describe the event in the present tense i.e instead of "going to" or "will be", they use
      "am" or "at"? (Yes or No)

"""
\caption{Prompt template for evaluation by checklist}
\label{evaluation}
\end{figure*}

\begin{figure*}[ht]
prompt =

    """
    You are comparing a generated response to an expert response in a given context. Here is the data:
    
    [BEGIN DATA]
    
    ************
    
    [User]: In about 1 month, I am horseback riding with my best friend on
    a farm in the fall. I am very happy
    
    ************
    
    [Expert]: In about a month, I am horseback riding in the beautiful fall leaves. I am with my friend on a beautiful farm. I am riding my favorite horse. The trail is covered in Orange and yellow leaves, and they are falling all around us. We are very happy and full of joy. It is nice and cool out.
    
    ************
    
    [Assistant]: In about 1 month, I am horseback riding with my best friend on a beautiful farm. The fall leaves are painting the landscape with warm colors. I am feeling incredibly happy, the joy bubbling up inside me as we ride together, laughing and chatting. The feeling of the horse moving beneath me is wonderful. It's a perfect day, and I am filled with a deep sense of contentment and joy.

    ************
    
    [END DATA]
    
    Compare the factual content of the generated response with the expert answer. Ignore any differences in style, grammar, or punctuation.
    Selecting any of the following options that apply:
    
    (A) The generated response contains all the same details (who? what? where? feeling?) as the expert answer.
    
    (B) There is a disagreement between the generated response and the expert answer.
    
    (C) The generated response describes a vivid event based on the definition of vividness as the expert answer

      Vivid definition: The text contains sufficient and specific details so that the event described is readily and easily imaginable. A considerable amount of context has been given regarding the event.
    
    (D) The generated response uses the correct format (i.e. start with a time frame, example, In about a month) as the expert answer?
    
    (E) The generated response describes the event in the present tense i.e instead of "going to" or "will be", they use
      "am" or "at" as the expert answer? 
      
  choice\_strings: ABCDE

    """
\caption{Prompt template for evaluation by comparison}
\label{fig6}
\end{figure*}

\textbf{Evaluation:}
As part of our formative evaluation, which has been aided by working with more than 20 participants who were willing to pilot-test the system, we have developed model-based approaches to assess the generated cues.
The LLM used for evaluation, and the LLM under assessment, do not necessarily need to be identical, although, for our application, they are the same (GPT-4).

In our specific application, the output of each interview is unique to the individual.
Accordingly, our assessment must consider the personalized nature of the generated output from these interviews, which shows a  rich diversity.
%, it's crucial to adopt the aforementioned technique.
Fortunately, we can use LLM-associated API calls to evaluate our chatbot output based on our current evaluation framework, which employs a checklist, and also is based on comparison\footnote{https://learn.deeplearning.ai/chatgpt-building-system}. Classification is another technique that we leverage for assessment. 

\textit{By Checklist:} The current assessment of the interview-guided EFT generation involves a checklist of requirements \cite{brown2022putting}.
A researcher checks if the participant uses the correct format, chooses a vivid event, and describes the event in the present tense \cite{brown2022putting}. These are three criteria that are considered to determine whether an EFT cue is good.
We can automate this assessment process by prompting the GPT-4 model to assess the generated EFT by the chatbot based on these criteria. The prompt template is given in Fig. \ref{evaluation}\footnote{https://github.com/openai/evals/tree/main/evals/registry/modelgraded}. As shown, the model is provided with a definition of vividness. The model outputs Yes or No for each of the three questions. The personalized EFTs generated by the chatbot have also been checked by a human, who has confirmed 
%that it is in consistent with 
the correctness of
the GPT-4 outputs regarding this assessment method.

\textit{By Comparison:} Another evaluation approach is to include a comparison between the generated EFT and an ideal answer\footnote{https://github.com/openai/evals/blob/main/evals/registry/modelgraded/fact.yaml}.
The prompt template shown in Fig. \ref{fig6} illustrates this approach. It originates from the OpenAI Evals project and is modified to incorporate the checklist and comparison techniques together. The example in Fig. 6 is considered an ideal answer based on the vividness for a participant who went through the interview-guided method. The chatbot is used by that participant to generate an EFT and then a comparison is performed by GPT-4. This can be applied in experimental designs where some participants work first with a researcher and then with the chatbot, while other participants work first with the chatbot and then with a researcher.
The ordering can be varied randomly across a set of participants, to account for learning effects. This would allow for determining how well the chatbot-generated EFT cue compares with a cue generated as the result of an interview with a researcher.

The model outputs any of the ABCDE options if the generated EFT satisfies the corresponding criterion. Since the model is prompted to write EFTs based on the details provided by the user for the event, option A checks if those details are included in the model output. Option B checks 
%the scenario when the generated EFT is irrelevant to the user's input and the expert response.
whether the generated EFT and the expert response are in agreement, with the chatbot following instructions and giving relevant content.
Similar checks are made regarding the other options.

\textit{By Classification}: Our previous work \cite{sara} examines content characteristics of EFT data generated by individuals suffering from conditions such as diabetes.
 We built classifiers to predict the content and characteristics of the
generated EFTs which can be used for in-depth analysis to pinpoint which text features contribute to positive health
results. The classifiers are based on prompting an LLM utilizing zero-shot and few-shot learning. One classification category is vividness, in which the classifier predicts
%the level of it. 
its level or degree.
Since vividness of events is one of the desired characteristics of EFT cues, one confirmation of the quality of the current chatbot is that this classifier returns `highly vivid.'
%The classifier prediction is highly vivid for the generated EFTs by the chatbot. vividness of events is one of the desired characteristics of EFT cues. 
Another classifier predicts the emotional tone, i.e., positive, neutral, or negative. The emotional tone prediction for the chatbot generated EFTs is positive, which confirms another important characteristic for the EFTs.

\section{Conclusion}
Taking into account the efficacy of Episodic Future Thinking (EFT) in guiding individuals toward healthier behaviors, this study explores the integration of AI chatbots into EFT cue generation. Traditional EFT cue generation involves humans conducting interviews to guide participants through cue generation.
%With the intent of facilitating the interview process and improving the quality of EFTs, 
To improve the quality of EFTs and faciliate the interview process,
our AI chatbot assists users in crafting personalized EFT cues. This could significantly enhance the efficiency and accessibility of EFT interventions, particularly for those facing educational or communication barriers.

The advances in Large Language Models (LLMs)  have revolutionized the use of chatbots and human-AI interaction.  Cutting-edge models like GPT-4 demonstrate human-level conversational performance, greatly attributed to innovative methods such as RLHF. The potency of these models is harnessed by prompt engineering, enabling customized interactions. Meanwhile, protective measures similar to the OpenAI Moderation API guarantee their responsible and safe deployment. Further, regarding formative evaluation, GPT-4 supports our checklist and comparison approaches. Additional quality confirmation is given through our classification framework, and pilot testing with more than 20 users.
%the two approaches as well as the classification framework to automatic evaluation, which supplements our pilot testing with more than 20 users.
\section{Future Work}
Work on the chatbot is ongoing. For example, one plan is to expand the classification approach for evaluation, initially as part of our formative evaluation, and later, during cue generation as an interim evaluation to improve quality. This will build on our recent research to classify cues according to 15 different characteristics \cite{sara}.
Separate research is exploring how each of the characteristics (e.g., about a celebration, about family, episodicity, etc.) of a cue correlates with improved health outcomes.
When that research determines which characteristics are most important for health, we can use our classifiers for such characteristics to help assess the quality of a chatbot-generated cue.
Then the chatbot can further engage a participant in editing their cue to include more helpful characteristics. The EFT research team is prepared to help with further formative evaluation, controlled user testing, and possible deployment in a clinical trial.
\section*{Acknowledgment}
Thanks go to people who have helped with pilot testing, and the NIH-funded EFT team. The EFT data is supported by  NIH NIDDK 3R01DK129567-02S1.

\bibliographystyle{unsrt}  

%\begin{thebibliography}{1}
\bibliography{references.bib}

% \end{thebibliography}

\end{document}